\documentclass[prd,showpacs,superscriptaddress,twocolumn,
floatfix,preprintnumbers,altaffilletter]{revtex4}
\usepackage{graphicx,url,amssymb,amsmath,longtable,rotating,color,units, 
wasysym,subfigure,epsfig,multirow}

\newcommand{\stochtrack}{{\em stochtrack}}
\newcommand{\burstegard}{{\em burstegard}}
\newcommand{\burstcluster}{{\em burstcluster}}

\begin{document}

\title{Searching for gravitational-wave transients with a qualitative signal model: seedless clustering strategies}


\author{Eric~Thrane}
\email{ethrane@ligo.caltech.edu}
\affiliation{LIGO Laboratory, California Institute of Technology, MS 100-36,
Pasadena, CA, 91125, USA}

\author{Michael Coughlin}
\affiliation{University of Cambridge, Cambridge, CB2 1TN, United Kingdom}

\date{\today}

\begin{abstract}
  Gravitational-wave bursts are observable as bright clusters of pixels in spectrograms of strain power.
  Clustering algorithms can be used to identify candidate gravitational-wave events.
  Clusters are often identified by grouping together seed pixels in which the power exceeds some threshold.
  If the gravitational-wave signal is long-lived, however, the excess power may be spread out over many pixels, none of which are bright enough to become seeds.
  Without seeds, the problem of detection through clustering becomes more complicated.
  In this paper we investigate seedless clustering algorithms in searches for long-lived narrowband gravitational-wave bursts.
  Using four astrophysically motivated test waveforms, we compare a seedless clustering algorithm to two algorithms using seeds.
  We find that the seedless algorithm can detect gravitational-wave signals (at fixed false-alarm and false-dismissal rate) at distances between $150$--$200\%$ greater than those achieved with the seed-based clustering algorithms, corresponding to significantly increased detection volumes: $420$--$740\%$.
  This improvement in sensitivity may extend the reach of second-generation detectors such as Advanced LIGO and Advanced Virgo deeper into astrophysically interesting distances.
\end{abstract}

\maketitle

\section{Introduction}\label{intro}
Searches for gravitational-wave (GW) transients typically fall into two classes.
``Burst'' searches employ only minimal assumptions to target unmodeled or difficult-to-model GW sources.
Other GW sources, such as coalescing neutron stars / black holes, produce readily predictable waveforms, making it possible to carry out a near-optimal search with a matched filter template bank.
However, it is also possible to design a GW transient search in between these two opposite ends of the spectra, where some information about the signal model is known, but not enough to produce a reliable template bank.
In this paper we investigate the possibility of GW transient searches for which we have a qualitative signal model, focusing in particular on models predicting GW signals, which are long-lived $\gtrsim\unit[10]{s}$ and narrowband, but which are otherwise poorly constrained.

\begin{figure*}[hbtp!]
  \begin{tabular}{cc}
    \psfig{file=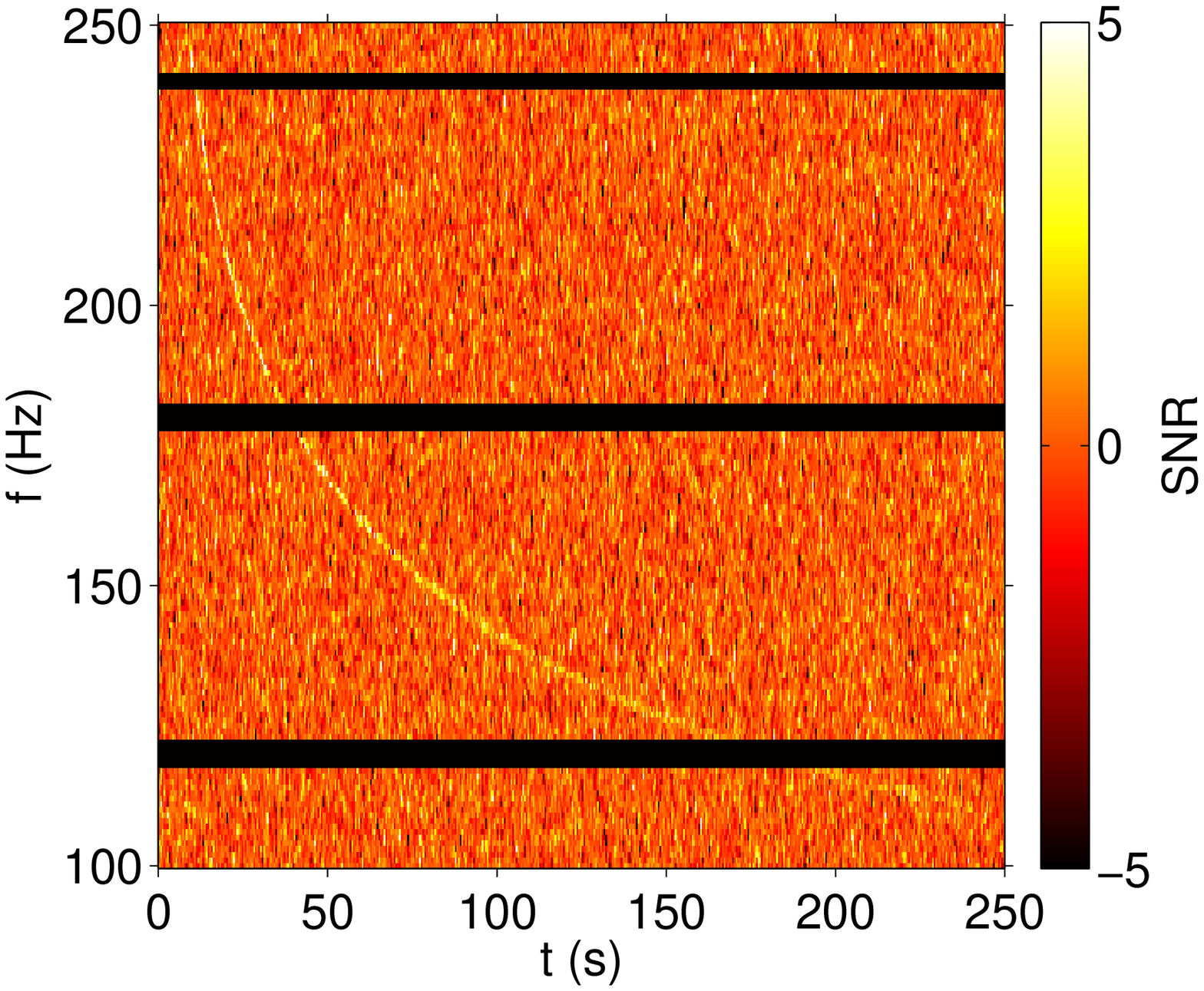,height=2.8in} &
    \psfig{file=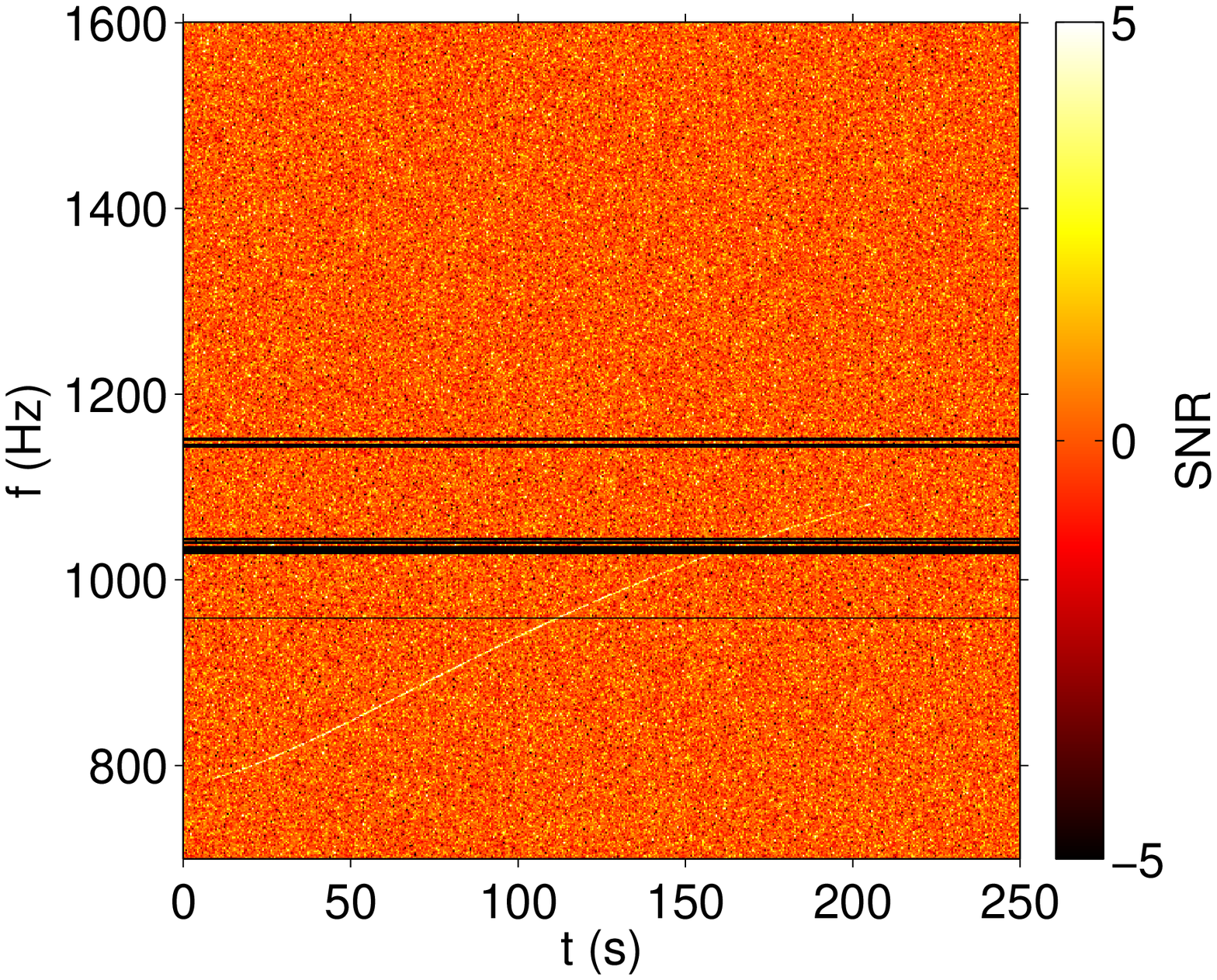,height=2.8in} \\
    \psfig{file=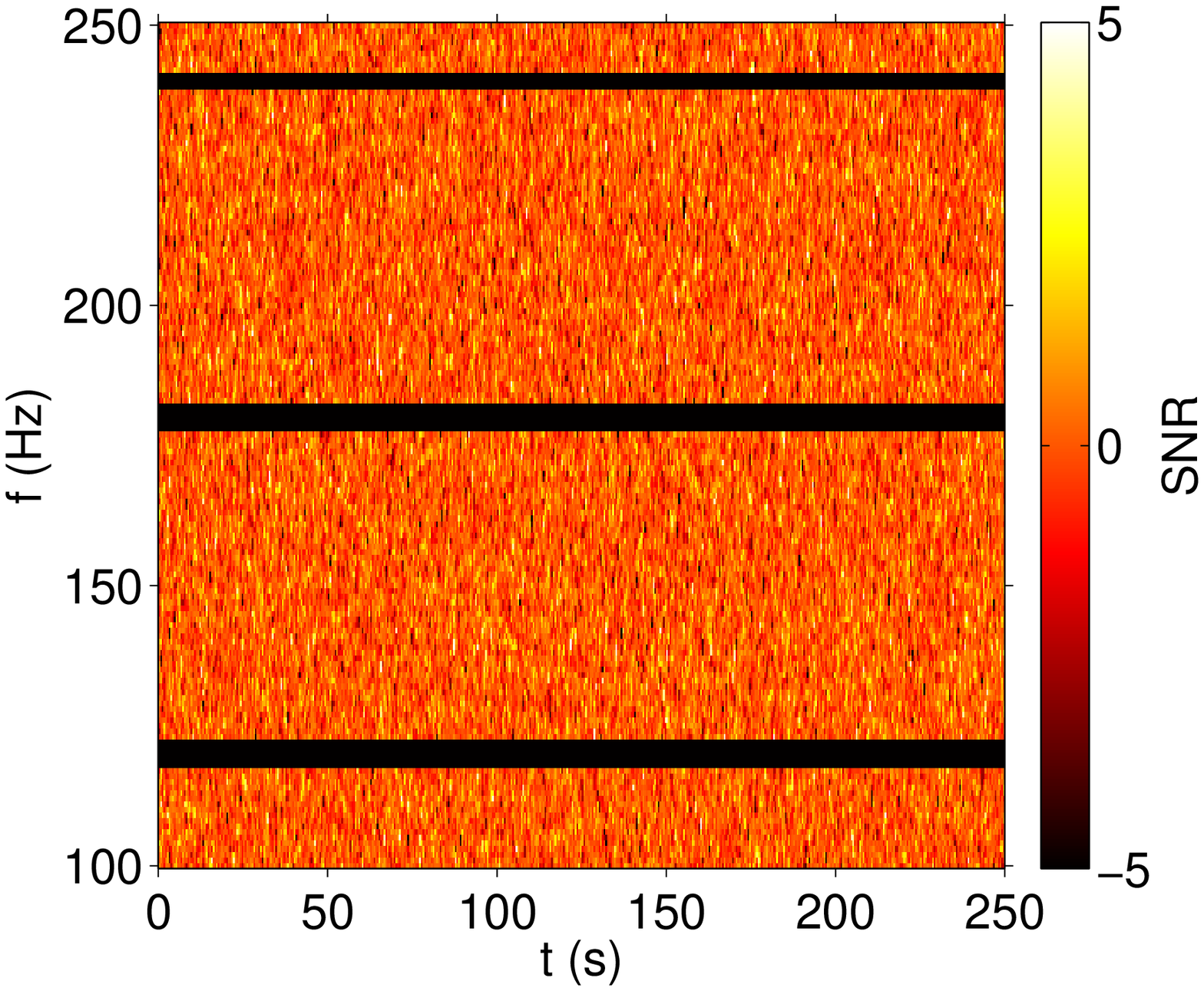,height=2.8in} &
    \psfig{file=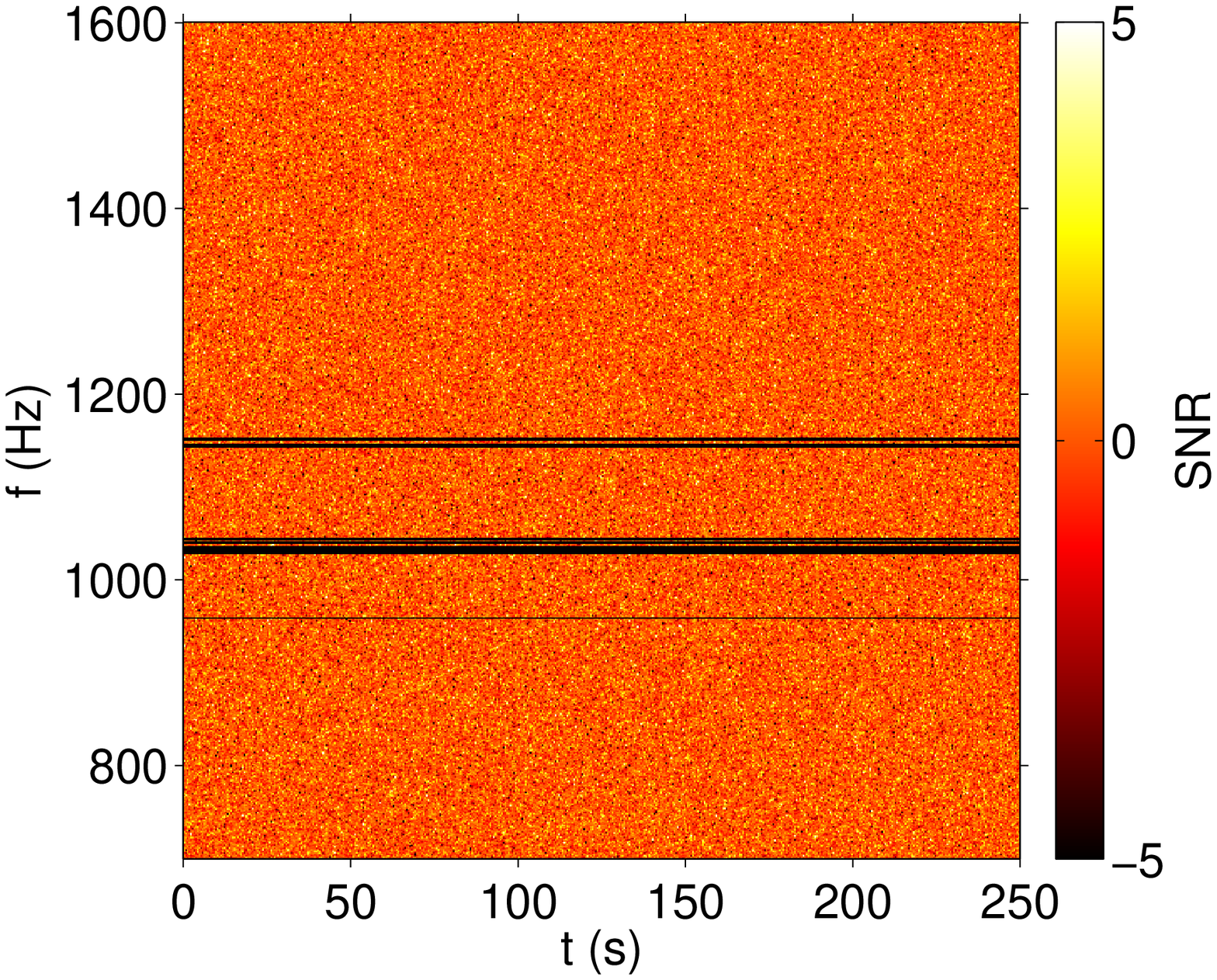,height=2.8in} \\
    \psfig{file=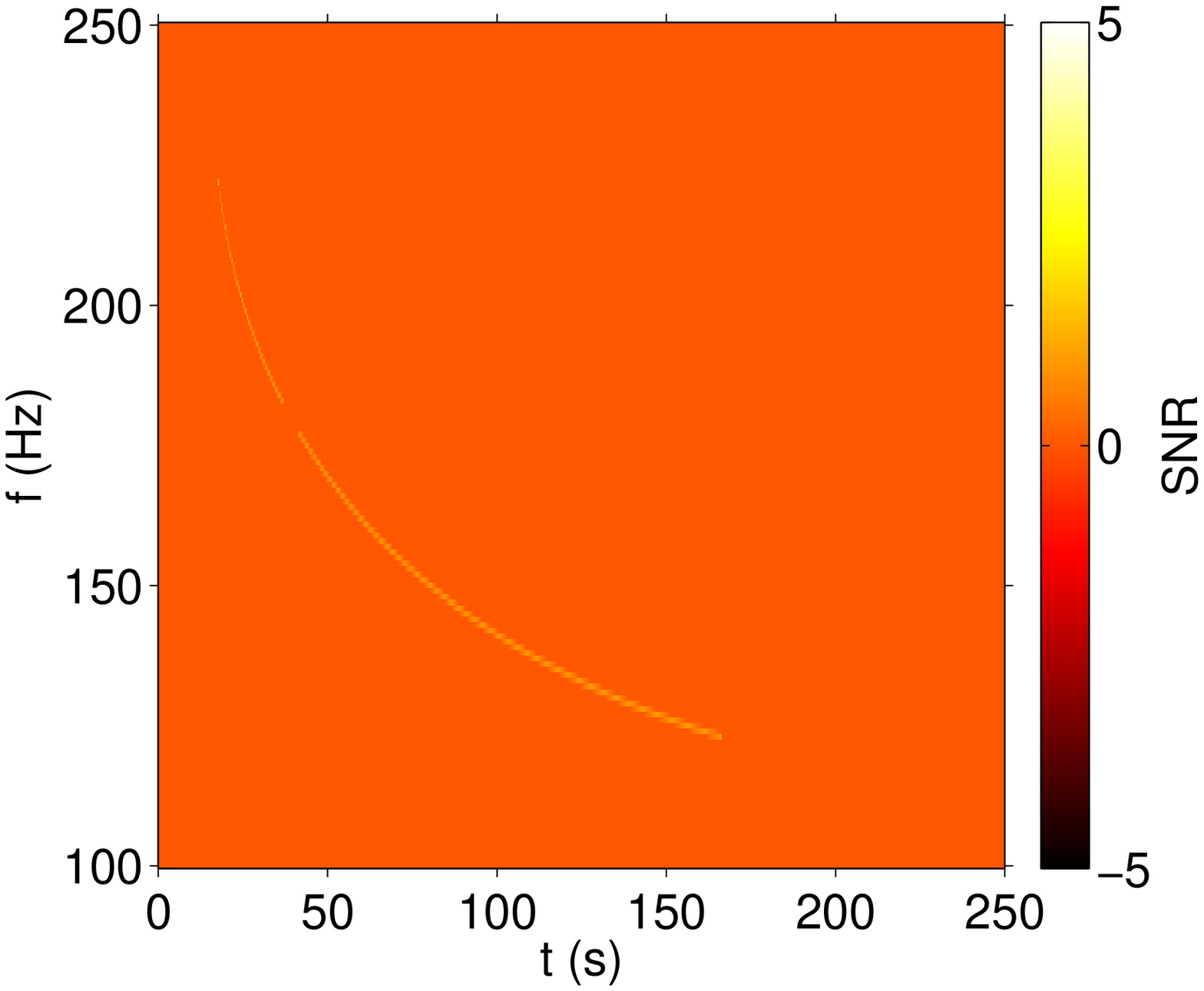,height=2.8in} &
    \psfig{file=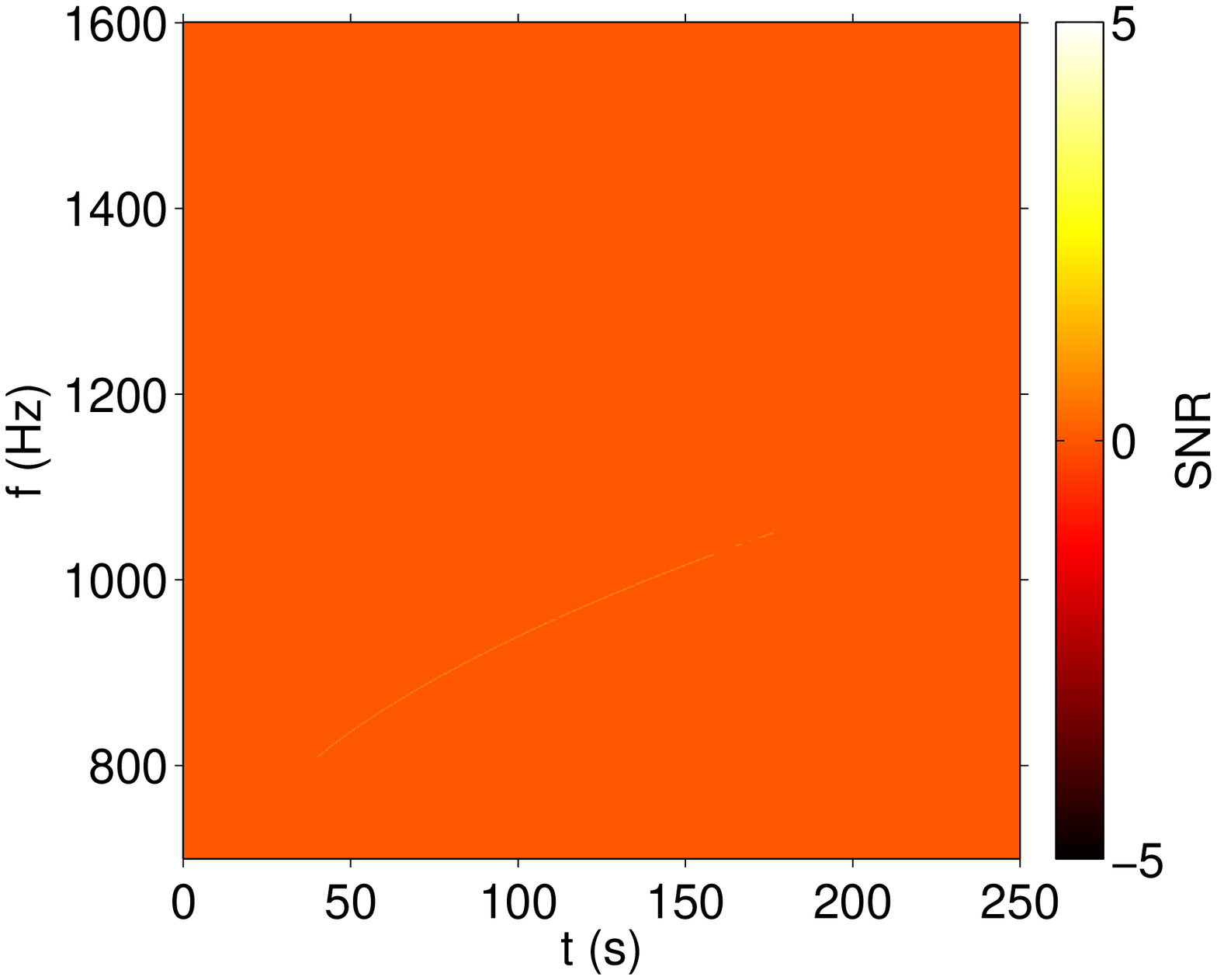,height=2.8in}
  \end{tabular}
  \caption{
    Injection recovery with seedless clustering using simulated Advanced LIGO noise.
    Top row: SNR spectrograms for relatively nearby signals.
    Left is a $d=\unit[150]{Mpc}$ accretion disk instability signal (ADI~2) and right is a $d=\unit[16]{Mpc}$ fallback accretion signal (FA~2); see Tab.~\ref{tab:waveforms}.
    The black horizontal lines are notches due to instrumental artifacts.
    Second row: the same as the first row, but the injected signals are further away, $d=\unit[360]{Mpc}$ (left) and $\unit[39]{Mpc}$ (right), and so the SNRs are less by $\approx6\times$.
    The tracks are all but invisible to the naked eye.
    Bottom row: the loudest recovered tracks obtained by analyzing the second-row spectrograms with stochtrack ($T=2\times10^8$ trials).
    Both clusters have $\text{FAP}<0.1\%$.
  }
  \label{fig:stochtrack}
\end{figure*}

Long-lived narrowband GW transients have been proposed to originate in a variety of astrophysical processes, most notably, in newborn neutron stars~\cite{pirothrane12,piro:11,piro:07,corsi} and black hole accretion disks following stellar collapse~\cite{kiuchi,vanputten:01,vanputten:08}.
Long-lived GW transients can be observed with excess strain power algorithms~\cite{stamp,pirothrane12}.
Signals show up as curved tracks on $ft$-maps (spectrograms) of strain power, see~Fig.\ref{fig:stochtrack}.

A number of clustering algorithms have been proposed to identify statistically significant GW signatures in strain power spectrograms, see, e.g.,~\cite{burstcluster,raffai,stamp,burstegard,waveburst,sylvestre}.
Most existing algorithms rely on the use of seeds: spectrogram pixels with excess power above some threshold~\footnote{
One exception is~\cite{murphy_sgr}, which describes a method for targeting quasimonochromatic signals.
By assuming monochromaticity, the space of possible waveforms is dramatically reduced, eliminating the need for seeds.
A second exception is the ``box search'' described in~\cite{stamp}, which avoids the use seeds by assuming the signal is broadband.
}.
The idea behind seed-based algorithms is that sufficiently loud GW signals induce excess power, which leads to the creation of seeds along a spectrogram track.
The clustering algorithm connects neighboring seeds in order to form a cluster.
(Different algorithms use different rules for connecting seeds.)
Next, the clustered seeds are combined to produce a detection statistic, which is used to determine if the cluster is consistent with detector noise.

One of the advantages of seed-based clustering is that only minimal assumptions need be made about the signal.
While different clustering rules may be better or worse for different signal models---e.g., narrowband tracks versus broadband blobs---most seed-based clustering algorithms can effectively cluster signals with arbitrary spectrographic morphology given a sufficiently high signal-to-noise ratio.

One disadvantage of seed-based clustering is that the signal must be loud enough to create seeds in the first place or the whole enterprise is doomed.
As we seek to study longer and weaker GW signals, this becomes increasingly problematic.
For a fixed energy budget, the average excess power in each of $N$ spectrogram pixels scales like $1/N$.
In other words, long signals are less likely to induce seeds than short signals, all else equal.

Here we investigate seedless clustering algorithms designed to target long and weak signals.
We propose a seedless clustering algorithm that will enforce additional assumptions about the signal model: that it is long-lived and narrowband.
By making these assumptions, we sacrifice some of the flexibility of seed-based clustering algorithms for improved sensitivity to a specific class of signal models.

The remainder of this paper is organized as follows.
In Section~\ref{formalism} we formulate the problem of detecting long-lived narrowband GW transients as a pattern recognition problem: how to detect tracks from GWs in strain power spectrograms.
In Section~\ref{clustering} we describe clustering algorithms with seeds and introduce an alternative seedless clustering algorithm.
In Section~\ref{results}, we describe a Monte Carlo study comparing the sensitivity of seed-based and seedless clustering algorithms.
We also repeat the analysis using recolored initial LIGO noise (with an unphysical time shift) to study whether the Monte Carlo results hold for non-ideal detector noise.
Finally, in Section~\ref{conclusions}, we summarize our findings, describe the limitations of our proposed algorithm, and discuss possibilities for future work.

\section{Formalism}\label{formalism}
Searches for long GW transients can be cast as pattern recognition problems~\cite{stamp}.
Strain time series $s_I(t')$ for detector $I$ are divided into segments of duration $\delta t$ with start times $t$.
(Note that $t'$ denotes sampling times whereas $t$ denotes segment start times.)
The Fourier transform of the detector-$I$ strain data in segment $t$ is denoted $\tilde{s}_I(t;f)$.

Following~\cite{stamp}, we define an estimator for strain cross-power in the $IJ$ detector pair $\hat{Y}_{IJ}(t;f)$ with associated variance $\hat\sigma^2(t;f)$:
\begin{equation}
  \begin{split}
    \hat{Y}(t;f) = \frac{2}{\cal N} \text{Re} \left[ 
      Q_{IJ}(t;f|\hat\Omega) \, \tilde{s}^\star_I(t;f)  \tilde{s}_J(t;f)
      \right] \\
    \hat\sigma^2(t;f) = \frac{1}{2} \left| Q_{IJ}(t;f|\hat\Omega) \right|^2
    P'_I(t;f) P'_J(t;f) .
  \end{split}
\end{equation}
Here ${\cal N}$ is an FFT normalization factor and $Q_{IJ}(t;f|\hat\Omega)$ is a filter function, which takes into account the relative time delays and the $IJ$ detector responses for a source located in the direction of $\hat\Omega$.
The filter function is defined such that $\hat{Y}(t;f)$ is an unbiased estimator for GW power~\cite{stamp}.
Meanwhile, $P'_I(t;f)$ and $P'_J(t;f)$ are the auto-power spectral densities for detectors $I$ and $J$ in the segments neighboring $t$.
For additional details, see~\cite{stamp}.

We define signal-to-noise ratio for a spectrogram pixel at $(t;f)$ as
\begin{equation}
  \rho(t;f)\equiv\hat{Y}(t;f)/\hat\sigma(t;f) .
\end{equation}
An array of $\rho(t;f)$ can be visualized as an $ft$-map as in Fig.~\ref{fig:stochtrack}.
Detector noise is distributed quasi-normally with mean $\langle\rho(t;f)\rangle=0$ while GW signals produce positive contributions to $\rho(t;f)$.
A loud long-lived narrowband transient, therefore, appears as a track of bright pixels in a spectrogram of $\rho(t;f)$.
If the GW signal is very weak, the track may not be visible by eye, though, there is still a statistical excess in $\rho(t;f)$ along the GW track.

The job of a clustering algorithm is to identify a cluster of pixels $\Gamma$, which, subject to some set of clustering rules, is more likely than any other cluster to be associated with a GW signal.
In order to determine which cluster among many is loudest, and in order to determine the statistical significance of a cluster, it is necessary to define a detection statistic characterizing the loudness of the entire cluster.
Following~\cite{stamp}, we define the cluster signal-to-noise ratio as
\begin{equation}\label{eq:sum}
  \text{SNR}_\text{tot} \equiv 
  \frac{
    \sum_{\left\{t;f\right\}\in\Gamma} w(t;f) \, Y(t;f)
  }{
    \left( \sum_{\left\{t;f\right\}\in\Gamma} w^2(t;f) \sigma^2(t;f) \right)^{1/2} 
    } .
\end{equation}
Here $w(t;f)$ is a weight factor, which can be chosen to emphasize certain frequencies and times depending on the detector noise, the expected GW signal, both, or neither.

By performing many pseudo experiments with Monte Carlo or time-shifted detector noise, it is possible to measure the probability density function $p(\text{SNR}_\text{tot})$ from which we determine the threshold $\text{SNR}_\text{tot}^\text{th}$ required for a detection at fixed false-alarm probability (FAP):
\begin{equation}
  \int_0^{\text{SNR}_\text{tot}^\text{th}} 
  d(\text{SNR}_\text{tot}) \,
  p(\text{SNR}_\text{tot}) = 1-\text{FAP} .
\end{equation}

The sensitivity of a clustering algorithm to a specific source can be characterized by the distance to which it can detect the source with $\text{SNR}_\text{tot}\geq\text{SNR}_\text{tot}^\text{th}$ with fixed FAP and fixed false-dismissal probability (FDP).
In this paper we define detection distance $d_0$ as the distance at which a GW signal can be observed above threshold with $\text{FAP}=0.1\%$ and $\text{FDP}=50\%$.
Detection distance is always defined for a specific gravitational waveform (model), so below we present results for several models.

\section{Clustering}\label{clustering}
In this section we discuss how different clustering algorithms can be used to identify tracks of excess power in spectrograms of $\rho(t;f)$.

\subsection{Seed-based clustering}
The first step for any seed-based algorithm is to apply a threshold in order to identify seeds:
\begin{equation}
  \rho(t;f) > \rho_\text{th} .
\end{equation}
The threshold is a tunable parameter that can be chosen so as to maximize $d_0$.
If $\rho_\text{th}$ is too small, there will be many seeds due to noise fluctuations, which leads to many loud noise clusters, ultimately harming the sensitivity of the search.
In fact, if $\rho_\text{th}$ is made sufficiently small, the typical density of seed pixels will be so great that seeds from noise fluctuations will form a single large cluster spreading throughout the spectrogram.
On the other hand, if $\rho_\text{th}$ is too large, only very loud signals will create seeds.
We find empirically that $\rho_\text{th}\approx0.75$ maximizes $d_0$ for the seed-based clustering algorithms considered here.

Next, the seeds are combined to produce clusters.
There are myriad ways of clustering seeds.
Linear clustering algorithms (e.g.,~\cite{burstegard}) combine seeds that fall within a fixed distance of each other.
Density-based clustering algorithms (e.g.,~\cite{burstcluster}) require that the number of seeds per unit area exceeds some threshold in order to be joined.
The ``locust'' algorithm~\cite{raffai}, meanwhile, is a local wandering scheme in which the two most significant neighboring seeds in some box are connected iteratively until no more seeds are available to connect.
It is also possible to combine the seeds along predefined paths specified by polynomials using a Hough algorithm~\cite{raffai}.
In the comparison that follows, we employ a linear clustering algorithm~\cite{burstegard} and a density based algorithm~\cite{burstcluster}, both of which are in use in GW transient analyses~\cite{stamp_glitch,ligo_sgr}.

One advantage of seed-based clustering is that most implementations, as a rule of thumb, can be made to operate with relatively modest computational resources.
Reducing a large number of pixels in a $\rho(t;f)$ spectrogram to a handful of seeds simplifies the clustering problem.

One disadvantage of seed-based clustering is that the excess strain power from long signals is spread out over many pixels and may therefore fail to produce seeds.
Another disadvantage arises from the presence of instrumental noise lines present GW strain data; see Fig.~\ref{fig:stochtrack}.
Noise lines must be notched to avoid numerous clusters from non-stationary noise.
The notches, in turn, create gaps over which it may be difficult to join seeds.
In the next subsection, we show how seedless clustering can overcome both of these obstacles.

\subsection{Seedless clustering}
A seedless clustering algorithm does not apply a threshold to $\rho(t;f)$.
An example of a previously proposed seedless clustering algorithm is the Radon algorithm~\cite{stamp}, which integrates $\rho(t;f)$ along every possible straight line that can be drawn through $\rho(t;f)$.
There are a number of limitations associated with the Radon algorithm, which we pause to study in order so that we might illuminate the path to a more effective clustering strategy.

First, the Radon algorithm assumes the track is well-described as a straight line in $ft$-space, which is a poor approximation for many realistic signals, see Fig.~\ref{fig:stochtrack}.
Second, it assumes that the signal persists for the duration of the spectrogram (or until the line intersects the top/bottom edges).
Finally, the background is needlessly increased by including nearly vertical lines, corresponding to short times, which do not conform to the assumed long-lived signal model.

We endeavor to address these shortcomings with a new seedless algorithm, which we call \stochtrack.
The basic idea of \stochtrack\ is to integrate $\rho(t;f)$ along monotonic $f(t)$ curves with arbitrary start and stop times subject to the constraint that the total duration is at least $t_\text{min}$ taken here to be $20$--$\unit[100]{s}$ depending on the model.
By allowing for curved tracks, we aim to better fit plausible GW signals.

The algorithm works as follows:
\begin{enumerate}
  \item Choose a random triplet of start-time, mid-time, and stop-time $(t_\text{start}, t_\text{mid}, t_\text{stop})$ such that $(t_\text{stop}-t_\text{start})\geq t_\text{min}$ and $t_\text{start}<t_\text{mid}<t_\text{end}$.
  \item Choose a random triplet of start-frequency, mid-frequency, and stop-frequency $(f_\text{start}, f_\text{mid}, f_\text{stop})$ such that $f_\text{start}\leq f_\text{mid}\leq f_\text{end}$ (up-chirping) or $f_\text{start}\geq f_\text{mid}\geq f_\text{end}$ (down-chirping).
  \item These two triplets correspond to three ordered pairs of $(f,t)$.
    Using the three ordered pairs as control points, form a quadratic B\'ezier curve~\cite{bezier} denoted $\Gamma$.
    (Other curve parameterizations, such as a cubic spline, are possible as well.)
  \item Following Eq.~\ref{eq:sum}, perform a weighted sum of the values of $\rho(t;f)$ in $\Gamma$ to calculate $\text{SNR}_\text{tot}$.
  \item Repeat the previous steps $T$ times.
    Record the cluster with the largest value of $\text{SNR}_\text{tot}$.
\end{enumerate}
Above we have described the \stochtrack\ algorithm in terms of a for-loop, but in practice it can be more computationally efficient to work with $T$-dimensional vectors of ordered pairs: $(\vec{t}_\text{start},\vec{f}_\text{start})$, $(\vec{t}_\text{mid},\vec{f}_\text{mid})$, and $(\vec{t}_\text{end},\vec{f}_\text{end})$.

In order to explore some of the computational subtleties of this calculation, it is worthwhile to consider a concrete example.
Consider a $\unit[151]{Hz}\times\unit[250]{s}$ spectrogram (as used below in Section~\ref{results}), which corresponds to $M\times N\equiv 151\times500$ pixels (see Fig.~\ref{fig:stochtrack}).
For these map dimensions, and assuming $t_\text{min}=\unit[100]{s}$, there are $\approx 2\times10^{13}$ possible combinations of ordered pairs making an exhaustive search unfeasible (see Appendix~\ref{computing}).
However, below we demonstrate that $T=2\times10^{7}$ random trials provides sufficient sampling to yield remarkable sensitivity gains with reasonable computational requirements.

Since the \stochtrack\ algorithm does not depend on the nearness of seed pixels, it is well-suited for realistic data with instrumental notches (see Fig.~\ref{fig:stochtrack}).
It is unaffected by the gaps in $\rho(t;f)$.

By design, the \stochtrack\ algorithm assumes a particular signal form.
Namely, the track is assumed to be reasonably well described by a quadratic B\'ezier curve with a duration of at least $t_\text{min}$.
(This family of signals includes as a subset all monochromatic tracks with duration of at least $t_\text{min}$.)
In reality, however, the quadratic B\'ezier curve will be only an approximate fit for an arbitrary monotonic curve.
Broadband signals and non-monotonic signals may be poorly fit.

\section{Comparison}\label{results}
In order to demonstrate the \stochtrack\ algorithm and compare it to seed-based clustering algorithms we perform a Monte Carlo study.
First, we generate Gaussian detector noise following the design sensitivity of Advanced LIGO (aLIGO) at high-power and zero-detuning~\cite{aligo}.
Using this simulated noise, we construct spectrograms of $\rho(t;f)$.
We analyze each spectrogram with three clustering algorithms: a linear clustering algorithm called \burstegard~\cite{burstegard}, a density-based clustering algorithm called \burstcluster~\cite{burstcluster}, and \stochtrack.
We run two versions of stochtrack: a default version with $T=2\times10^{7}$ trials and a computationally more expensive deep-search version with $2\times10^{8}$ trials denoted ``\stochtrack~$10\times$.''
By running both the default \stochtrack\ and \stochtrack~$10\times$, we investigate how detection distance scales with the number of trials.

For each algorithm, we determine the threshold $\text{SNR}_\text{tot}^\text{th}$ corresponding to $\text{FAP}=0.1\%$ (see Section~\ref{formalism}).
Once we have obtained the thresholds, we perform additional Monte Carlo studies in which a signal is added to the simulated noise.
By looping over a range of source distances, we can vary the signal strength, and determine the $\text{FAP}=0.1\%$, $\text{FDP}=50\%$ detection distance $d_0$ for each algorithm; see Section~\ref{formalism}.
We consider four toy-model waveforms: two down-chirping accretion-disk instability (ADI) waveforms inspired by~\cite{vanputten:01,vanputten:08} and calculated following~\cite{lucia} and two up-chirping fallback accretion (FA) powered waveforms from~\cite{pirothrane12}; see Table~\ref{tab:waveforms}.

In the FA model, a newborn neutron star is spun up through fallback accretion following a supernova~\cite{piro:11,pirothrane12}.
The neutron star undergoes a dynamical or secular instability, which induces a time varying quadrupole moment, which in turn powers the emission of narrowband GWs until a black hole is formed and the signal is cut off.
In the ADI model, clumps form in the accretion disk surrounding a black hole formed following stellar collapse~\cite{vanputten:01,vanputten:08,lucia}.
The motion of the clumps leads to the emission of narrowband GWs.
The ADI waveforms are normalized to assume a GW energy budget of $E_\text{GW}=0.1 M_\odot$~\cite{vanputten:01}.

The durations and frequency range of each waveform are given in Table~\ref{tab:waveforms}.
The waveform parameters are listed in the Appendix~\ref{parameters}.
The ADI waveforms are analyzed in a band between $100$--$\unit[250]{Hz}$ while the FA waveforms are analyzed in a band between $700$--$\unit[1600]{Hz}$.

For our present purposes, we work under the assumption that the GW source location is known, e.g., from an electromagnetic trigger such as a gamma-ray burst or a supernova.
We further assume that the time of GW emission is constrained to a small $\unit[250]{s}$ ``on-source'' window.
While the $\unit[250]{s}$ window size is comparable to some previous triggered searches for GW bursts, e.g.,~\cite{grb:s5vsr1}, there are many signal models that would require a significantly larger on-source region~\cite{corsi,pirothrane12,vanputten:01,vanputten:08,stamp}.
Despite this, we restrict the on-source window to $\unit[250]{s}$ in order to compare different clustering algorithms with a limited computational cost.
It is possible to extend this type of analysis to study a larger on-source region at increased computational cost (or with diminished sensitivity at the same computational cost).

We assume that each source is optimally oriented (face-on), which is a reasonable assumption if we assume that the search is carried out following a highly-beamed electromagnetic trigger.
We further assume that the detectors are optimally aligned to achieve the maximal possible signal-to-noise ratio.
The detection distance obtained by averaging over detector orientations is $\approx60\%$ the value obtained by assuming optimal-aligned distanceç.

\begin{table}
  \begin{tabular}{|c|c|c|c|c|}
    \hline
    waveform & duration (s) & $f_\text{min}$--$f_\text{max}$ (Hz) & 
    $\delta{t}\times\delta{f}$ & $t_\text{min}$ \\\hline
    ADI 1 & $39$ & $130$--$170$ & 
    $\unit[1]{s}\times\unit[1]{Hz}$ & $\unit[35]{s}$ \\\hline
    ADI 2 & $230$ & $110$--$260$ & 
    $\unit[1]{s}\times\unit[1]{Hz}$ & $\unit[100]{s}$ \\\hline
    FA 1 & $25$ & $1170$--$1530$ & 
    $\unit[0.5]{s}\times\unit[2]{Hz}$ & $\unit[20]{s}$ \\\hline
    FA 2 & $200$ & $790$--$1080$ & 
    $\unit[1]{s}\times\unit[1]{Hz}$ & $\unit[100]{s}$ \\\hline
  \end{tabular}
  \caption{
    A summary of the waveforms used in our Monte Carlo study.
    The second and third columns describe the duration and frequency range of the waveform respectively.
    The fourth column gives the spectrogram resolution used to analyze each waveform.
    The fifth column specifies the minimum signal duration assumed in each search.
    The ADI waveforms are down-chirping accretion-disk instability waveforms~\cite{vanputten:01,vanputten:08,lucia} while the FA waveforms are up-chirping fallback accretion powered waveforms~\cite{pirothrane12,piro:11}.
  }
  \label{tab:waveforms}
\end{table}

The results are summarized in Table~\ref{tab:distances}.
We find that, depending on the waveform, the default \stochtrack\ improves on the seed-based clustering algorithms by a factor ranging from $150$--$180\%$ in distance, or equivalently, $320$--$560\%$ in volume.
For \stochtrack~$10\times$, the improvement is $160$--$200\%$ in distance, or equivalently, $420$--$740\%$ in volume.

\begin{table}
  \begin{tabular}{|c|c|c|c|c|c|}
    \hline
    \multicolumn{1}{|c|}{waveform} &
    \multicolumn{1}{|c|}{algorithm} &
    \multicolumn{2}{c|}{distance} &
    \multicolumn{1}{c|}{volume} \\
    & & absolute & \% & \% \\\hline
    \multirow{4}{*}{ADI 1}
    & \burstcluster\ & $\unit[330]{Mpc}$ & $90$ & $74$  \\
    & \burstegard\ & $\unit[370]{Mpc}$ & $100$ & $100$ \\
    & \stochtrack\ & $\unit[540]{Mpc}$ & $150$ & $320$ \\
    & \stochtrack\ $10\times$ & $\unit[590ó]{Mpc}$ & $160$ & $420$ \\\hline
    \multirow{4}{*}{ADI 2}
    & \burstcluster\ & $\unit[170]{Mpc}$ & $91$ & $76$  \\
    & \burstegard\ & $\unit[190]{Mpc}$ & $100$ & $100$ \\
    & \stochtrack\ & $\unit[340]{Mpc}$ & $180$ & $560$ \\
    & \stochtrack\ $10\times$ & $\unit[370]{Mpc}$ & $200$ & $740$ \\\hline
    \multirow{3}{*}{FA 1}
    & \burstegard\ & $\unit[17]{Mpc}$ & $100$ & $100$ \\
    & \stochtrack\ &  $\unit[29]{Mpc}$ & $150$ & $320$ \\
    & \stochtrack\ $10\times$ & $\unit[35]{Mpc}$ & $180$ & $560$ \\\hline
    \multirow{3}{*}{FA 2}
    & \burstegard\ & $\unit[25]{Mpc}$ & $100$ & $100$ \\
    & \stochtrack\ & $\unit[36]{Mpc}$ & $150$ & $320$ \\
    & \stochtrack\ $10\times$ & $\unit[40]{Mpc}$ & $160$ & $420$ \\\hline
  \end{tabular}
  \caption{
    A comparison of the sensitivity achieved with three different clustering algorithms using aLIGO Monte Carlo noise.
    {\em Burstcluster}~\cite{burstcluster} and \burstegard~\cite{burstegard} use seeds whereas \stochtrack\ is seedless.
    By default, \stochtrack\ performs $T=2\times10^7$ trials.
    We also report results for \stochtrack\ $10\times$ using $T=2\times10^8$ trials.
    (Note that \burstcluster\ distances are only available for the ADI waveforms since the algorithm is too slow without modification to analyze the larger FA spectrograms.)
    ``Distance'' refers to the distance at which a GW source can be detected with false alarm probability $=0.1\%$ and false dismissal probability $=50\%$.
    We list both the absolute distance in Mpc and the \% relative to the \burstegard\ algorithm.
    The ADI waveforms have been scaled assuming an energy budget of $E_\text{GW}=0.1 M_\odot$.
    Volume is given in \% relative to the \burstegard\ algorithm.
    \label{tab:distances}
  }
\end{table}

For bright extra-galactic ADI signals with $E_\text{GW}=0.1 M_\odot$~\cite{vanputten:01,vanputten:08,lucia}, we obtain \stochtrack~$10\times$ detection distances of $d_0=$370--$\unit[590]{Mpc}$.
The rate of gamma-ray bursts within this distance range is $\sim0.1$--$\unit[1]{year^{-1}}$, which suggests that seedless clustering could facilitate the detection of an ADI-type signal by aLIGO~\cite{aligo} / aVirgo~\cite{virgo}.

For FA sources~\cite{piro:11,pirothrane12}, we obtain \stochtrack~$10\times$ detection distances of $d_0=35$--$\unit[40]{Mpc}$~\footnote{
  The baseline detection distances for the seed-based \burstegard\ clustering algorithm presented here are slightly larger compared to the distances stated in~\cite{pirothrane12}.
  There are some differences between the two calculations; e.g., the spectrograms are different sizes and here we use a more optimal spectrogram resolution.
  Most of the discrepancy, however, is accounted for by the use here of a higher-sensitivity aLIGO noise curve.
}.
The rate of supernovae in this volume is sufficiently high that aLIGO and aVirgo can expect a promising electromagnetic trigger rate of $\gtrsim\unit[1]{year^{-1}}$~\cite{pirothrane12}.

The gain in sensitivity is not without added computational cost.
On a currently typical computer, the \burstegard\ algorithm is capable of analyzing a single $151\times500$ pixel spectrogram in just $\unit[1.3]{s}$ while the default \stochtrack\ algorithm takes $\unit[1100]{s}$ ($\unit[18]{min}$) to analyze the same data.
The \stochtrack\ computation time scales linearly with the number of trials.
By increasing the number of trials by a factor of ten, it is possible to increase the detection distance by $\approx10\%$, but the computation time grows to $\unit[1.1\times10^{4}]{s}$ ($\approx\unit[3]{hr}$).

While $\approx\unit[3]{hr}$ of computing time is not especially burdensome in and of itself, an actual observational analysis will require many ($\gtrsim100$) pseudo-experiments with time-shifted data.
If we further assume that the algorithm is applied to $\approx50$ triggers (for example, from gamma-ray bursts), using an on-source region that is larger than the one used here by a factor of $\approx50$, then the estimated computing time is nine weeks on 500 dedicated nodes.

The number of trials can be tuned to match available computational resources.
In the event of a detection candidate, additional trials can be carried out to perform a more sensitive follow-up search.
Similarly, a seedless clustering algorithm such as \stochtrack\ could be used to follow up on candidates identified by a less sensitive, but computationally cheaper algorithm designed to look for untriggered GW transients in an all-sky, all-time search.

As an additional check, we repeat the comparison of clustering algorithms using initial LIGO noise~\footnote{The data are taken in between GPS times $822917487$--$847549782$.} recolored to match the aLIGO noise curve expected for zero-detuning and high laser power~\cite{aligo}.
This allows us to test the performance of the algorithm with non-stationary noise transients and other instrumental artifacts~\cite{stamp_glitch,lsc_glitch}.
An unphysical time shift is introduced between the two strain channels in order to remove any coherent signals.
The recolored noise results are summarized in Table~\ref{tab:recolored}.
The default \stochtrack\ improves on the seed-based clustering algorithms by a factor ranging from $150$--$180\%$ in distance, or equivalently, $320$--$560\%$ in volume.
For \stochtrack~$10\times$, the improvement is $160$--$200\%$ in distance, or equivalently, $420$--$740\%$ in volume.
The similarity between the Monte Carlo and recolored noise results is consistent with previous results~\cite{stamp_glitch} and suggests that the expected sensitivity gains from seedless clustering are not dependent on the assumption of idealized detector noise.

\begin{table}
  \begin{tabular}{|c|c|c|c|c|c|}
    \hline
    \multicolumn{1}{|c|}{waveform} &
    \multicolumn{1}{|c|}{algorithm} &
    \multicolumn{2}{c|}{distance} &
    \multicolumn{1}{c|}{volume} \\
    & & absolute & \% & \% \\\hline
    \multirow{4}{*}{ADI 1}
    & \burstcluster\ & $\unit[280]{Mpc}$ & $83$ & $57$  \\
    & \burstegard\ & $\unit[330]{Mpc}$ & $100$ & $100$ \\
    & \stochtrack\ & $\unit[540]{Mpc}$ & $160$ & $420$ \\
    & \stochtrack\ $10\times$ & $\unit[540]{Mpc}$ & $160$ & $420$ \\\hline
    \multirow{4}{*}{ADI 2}
    & \burstcluster\ & $\unit[159]{Mpc}$ & $91$ & $76$  \\
    & \burstegard\ & $\unit[170]{Mpc}$ & $100$ & $100$ \\
    & \stochtrack\ & $\unit[310]{Mpc}$ & $180$ & $560$ \\
    & \stochtrack\ $10\times$ & $\unit[340]{Mpc}$ & $200$ & $740$ \\\hline
    \multirow{3}{*}{FA 1}
    & \burstegard\ & $\unit[22]{Mpc}$ & $100$ & $100$ \\
    & \stochtrack\ &  $\unit[32]{Mpc}$ & $150$ & $320$ \\
    & \stochtrack\ $10\times$ & $\unit[35]{Mpc}$ & $160$ & $420$ \\\hline
    \multirow{3}{*}{FA 2}
    & \burstegard\ & $\unit[25]{Mpc}$ & $100$ & $100$ \\
    & \stochtrack\ & $\unit[40]{Mpc}$ & $160$ & $420$ \\
    & \stochtrack\ $10\times$ & $\unit[44]{Mpc}$ & $180$ & $560$ \\\hline
  \end{tabular}
  \caption{
    The same as Table~\ref{tab:distances} except we use recolored initial LIGO noise with an unphysical time shift instead of Monte Carlo.
    \label{tab:recolored}
  }
\end{table}

\section{conclusions}\label{conclusions}
Given a fixed energy budget, a long-lived GW transient produces less excess strain power at any given moment than a short burst.
Thus, a long-lived transient is less likely than a short burst with the same total available energy to produce the seed pixels necessary for many traditional clustering algorithms to recover a statistically significant signal.
In order to address this, we propose a seedless clustering algorithm called \stochtrack\ designed to detect signals too weak to produce seeds.
We apply \stochtrack\ to several long-lived narrowband signal models and find that it significantly improves detectability compared to two benchmark clustering algorithms, both of which use seeds.

There are a number of ways in which it might be possible to improve the \stochtrack\ algorithm.
In our current implementation, tracks are fit approximately with quadratic B\'ezier curves.
It may be possible to achieve further improvements in sensitivity using a different, more flexible curve parameterization.
The trick with any new parameterization is to better fit test waveforms without expanding the parameter space to the point where the increase in background offsets the gain in signal.

The algorithm may also benefit from improvements in computational efficiency.
A more efficient design and/or implementation might reduce the time required to analyze a spectrogram.
Reduced computation time, in turn, could facilitate deeper searches (with more trials) and/or searches with large on-source regions.
For example, it might be possible to replace the random track generation step with a deterministic process, which more intelligently samples the space of possible curves.
One can even imagine the creation of a template bank of curves analogous to the matched filter template banks used for compact binary coalescence searches.
(Unlike a matched filter template bank, a \stochtrack\ template bank would not contain phase information.)

An area of future research is the application of seedless clustering algorithms to the recovery of compact binary coalescence signals.
Of particular interest are regions of parameter space for which it is difficult to create matched filter template banks, e.g., systems with spin and/or eccentricity.

Cornish and Romano have recently emphasized the connection between data analysis algorithms and the signal model for which they are optimal~\cite{cornish_romano}.
Following the logic of~\cite{box} and~\cite{cornish_romano}, \stochtrack\ is an optimal search algorithm (in the limit that $T\rightarrow\infty$) for the class of signals described by quadratic B\'ezier curves in spectrograms of GW power with durations greater than $t_\text{min}$.
Given additional information about the signal model, a seedless clustering algorithm such as \stochtrack\ could be tuned appropriately to be more nearly optimal.

\begin{acknowledgments}
We thank Anthony Piro for sharing the fallback accretion waveforms used in this analysis.
We thank Shivaraj Kandhasamy and Nelson Christensen for helpful comments.
ET is a member of the LIGO Laboratory, supported by funding from United States National Science Foundation.
LIGO was constructed by the California Institute of Technology and Massachusetts Institute of Technology with funding from the National Science Foundation and operates under cooperative agreement PHY-0757058.
MC is supported by the Winston Churchill Foundation of the United States.
This paper has been assigned LIGO document number LIGO-P1300103.
\end{acknowledgments}

\begin{appendix}
  \section{Computational scaling}\label{computing}
We estimate the number of possible quadratic B\'ezier tracks with duration greater than $t_\text{min}$ in a $M\times N$ spectrogram.
For the sake of simplicity, we assume that $t_\text{min}$ is in units of time bins.
The number of frequency triplets is given by
\begin{equation}
  2 \int_0^M df_3 \int_0^{f_3} df_1 \int_{f_1}^{f_3} df_2 = \frac{M^3}{3} .
\end{equation}
Here $f_1$ is the start frequency, $f_2$ is the mid frequency, and $f_3$ is the end frequency.
The factor of $2$ comes from the fact that the signal can be both up-chirping or down-chirping.

The number of time triplets is given by
\begin{equation}
  \int_{t_\text{min}}^N dt_3 \int_0^{t_3-t_\text{min}} dt_1 \int_{t_1}^{t_3} dt_2 =
  \frac{N^3}{6} - \frac{t_\text{min}^2 N}{2} + \frac{t_\text{min}^3}{3} .
\end{equation}
Here $t_1$ is the start time, $t_2$ is the mid time, and $t_3$ is the end time.
Thus, the total number of possible tracks is
\begin{equation}
  \frac{M^3}{3} \left( 
  \frac{N^3}{6} - \frac{t_\text{min}^2 N}{2} + \frac{t_\text{min}^3}{3}
  \right) .
\end{equation}

\section{Model parameters}\label{parameters}
The FA waveforms~\cite{piro:11,pirothrane12} are parameterized by the initial protoneutron star mass $M_0$, the maximum neutron star mass $M_\text{max}$, a dimensionless factor characterizing the supernovae explosion energy $\eta\approx0.1$--$10$, and the protoneutron star radius $R_0$.
The two FA waveforms used here assume the following parameters:

\begin{table}[h]
  \begin{tabular}{|c|c|c|c|c|}
    \hline
    waveform & $M_0$ ($M_\odot$) & $M_\text{max}$ ($M_\odot$) & $\eta$ & $R_0$ (km) \\\hline
    FA 1 & $1.3$ & $2.5$ & $10$ & $20$ \\\hline
    FA 2 & $1.3$ & $2.5$ & $1$ & $25$ \\\hline
  \end{tabular}
  \caption{Parameters for FA waveforms.  See~\cite{pirothrane12} for additional details.}
\end{table}

The ADI waveforms~\cite{lucia} are parmeterized by black hole mass $M_\text{BH}$, dimensionless spin parameter $\alpha^\star=[0,1)$, the fraction of the accretion disk mask that forms clumps $\epsilon\approx0.01$--$0.2$, and the torus mass $m$.
The two ADI waveforms used here assume the following parameters:

\begin{table}[h]
  \begin{tabular}{|c|c|c|c|c|}
    \hline
    waveform & $M_\text{BH}$ ($M_\odot$) & $\alpha$ & $\epsilon$ & $m$ ($M_\odot$) \\\hline
    ADI 1 & $5$ & $0.3$ & $0.05$ & $1.5$ \\\hline
    ADI 2 & $10$ & $0.95$ & $0.04$ & $1.5$ \\\hline
  \end{tabular}
  \caption{Parameters for ADI waveforms.  See~\cite{lucia} for additional details.}
\end{table}

\end{appendix}

\bibliography{stochtrack}

\begin{thebibliography}{24}
\expandafter\ifx\csname natexlab\endcsname\relax\def\natexlab#1{#1}\fi
\expandafter\ifx\csname bibnamefont\endcsname\relax
  \def\bibnamefont#1{#1}\fi
\expandafter\ifx\csname bibfnamefont\endcsname\relax
  \def\bibfnamefont#1{#1}\fi
\expandafter\ifx\csname citenamefont\endcsname\relax
  \def\citenamefont#1{#1}\fi
\expandafter\ifx\csname url\endcsname\relax
  \def\url#1{\texttt{#1}}\fi
\expandafter\ifx\csname urlprefix\endcsname\relax\def\urlprefix{URL }\fi
\providecommand{\bibinfo}[2]{#2}
\providecommand{\eprint}[2][]{\url{#2}}

\bibitem[{\citenamefont{Piro and Thrane}(2012)}]{pirothrane12}
\bibinfo{author}{\bibfnamefont{A.~L.} \bibnamefont{Piro}} \bibnamefont{and}
  \bibinfo{author}{\bibfnamefont{E.}~\bibnamefont{Thrane}},
  \bibinfo{journal}{Astrophys. J.} \textbf{\bibinfo{volume}{761}},
  \bibinfo{pages}{63} (\bibinfo{year}{2012}).

\bibitem[{\citenamefont{Piro and Ott}(2011)}]{piro:11}
\bibinfo{author}{\bibfnamefont{A.~L.} \bibnamefont{Piro}} \bibnamefont{and}
  \bibinfo{author}{\bibfnamefont{C.~D.} \bibnamefont{Ott}},
  \bibinfo{journal}{Astrophys. J.} \textbf{\bibinfo{volume}{736}},
  \bibinfo{pages}{108} (\bibinfo{year}{2011}).

\bibitem[{\citenamefont{Piro and Pfahl}(2007)}]{piro:07}
\bibinfo{author}{\bibfnamefont{A.~L.} \bibnamefont{Piro}} \bibnamefont{and}
  \bibinfo{author}{\bibfnamefont{E.}~\bibnamefont{Pfahl}},
  \bibinfo{journal}{Astrophys. J.} \textbf{\bibinfo{volume}{658}},
  \bibinfo{pages}{1173} (\bibinfo{year}{2007}).

\bibitem[{\citenamefont{Corsi and M\'esz\'aros}(2009)}]{corsi}
\bibinfo{author}{\bibfnamefont{A.}~\bibnamefont{Corsi}} \bibnamefont{and}
  \bibinfo{author}{\bibfnamefont{P.}~\bibnamefont{M\'esz\'aros}},
  \bibinfo{journal}{Astrophys. J.} \textbf{\bibinfo{volume}{702}},
  \bibinfo{pages}{1171} (\bibinfo{year}{2009}).

\bibitem[{\citenamefont{Kiuchi et~al.}(2011)\citenamefont{Kiuchi, Shibata,
  Montero, and Font}}]{kiuchi}
\bibinfo{author}{\bibfnamefont{K.}~\bibnamefont{Kiuchi}},
  \bibinfo{author}{\bibfnamefont{M.}~\bibnamefont{Shibata}},
  \bibinfo{author}{\bibfnamefont{P.~J.} \bibnamefont{Montero}},
  \bibnamefont{and} \bibinfo{author}{\bibfnamefont{J.~A.} \bibnamefont{Font}},
  \bibinfo{journal}{Phys. Rev. Lett.} \textbf{\bibinfo{volume}{106}},
  \bibinfo{pages}{251102} (\bibinfo{year}{2011}).

\bibitem[{\citenamefont{van Putten}(2001)}]{vanputten:01}
\bibinfo{author}{\bibfnamefont{M.~H. P.~M.} \bibnamefont{van Putten}},
  \bibinfo{journal}{Phys. Rev. Lett.} \textbf{\bibinfo{volume}{87}},
  \bibinfo{pages}{091101} (\bibinfo{year}{2001}).

\bibitem[{\citenamefont{van Putten}(2008)}]{vanputten:08}
\bibinfo{author}{\bibfnamefont{M.~H. P.~M.} \bibnamefont{van Putten}},
  \bibinfo{journal}{Astrophys. J. Lett.} \textbf{\bibinfo{volume}{684}},
  \bibinfo{pages}{91} (\bibinfo{year}{2008}).

\bibitem[{\citenamefont{Thrane et~al.}(2011)\citenamefont{Thrane, Kandhasamy,
  Ott et~al.}}]{stamp}
\bibinfo{author}{\bibfnamefont{E.}~\bibnamefont{Thrane}},
  \bibinfo{author}{\bibfnamefont{S.}~\bibnamefont{Kandhasamy}},
  \bibinfo{author}{\bibfnamefont{C.~D.} \bibnamefont{Ott}},
  \bibnamefont{et~al.}, \bibinfo{journal}{Phys. Rev. D}
  \textbf{\bibinfo{volume}{83}}, \bibinfo{pages}{083004}
  (\bibinfo{year}{2011}).

\bibitem[{\citenamefont{Khan and Chatterji}(2009)}]{burstcluster}
\bibinfo{author}{\bibfnamefont{R.}~\bibnamefont{Khan}} \bibnamefont{and}
  \bibinfo{author}{\bibfnamefont{S.}~\bibnamefont{Chatterji}},
  \bibinfo{journal}{Classical Quantum Gravity} \textbf{\bibinfo{volume}{26}},
  \bibinfo{pages}{155009} (\bibinfo{year}{2009}).

\bibitem[{\citenamefont{Raffai et~al.}(2007)}]{raffai}
\bibinfo{author}{\bibfnamefont{P.}~\bibnamefont{Raffai}} \bibnamefont{et~al.},
  \bibinfo{journal}{Classical Quantum Gravity} \textbf{\bibinfo{volume}{24}},
  \bibinfo{pages}{S457} (\bibinfo{year}{2007}).

\bibitem[{\citenamefont{Prestegard and Thrane}(2012)}]{burstegard}
\bibinfo{author}{\bibfnamefont{T.}~\bibnamefont{Prestegard}} \bibnamefont{and}
  \bibinfo{author}{\bibfnamefont{E.}~\bibnamefont{Thrane}},
  \bibinfo{journal}{LIGO DCC} p. \bibinfo{pages}{L1200204}
  (\bibinfo{year}{2012}),
  \bibinfo{note}{\url{https://dcc.ligo.org/cgi-bin/DocDB/ShowDocument?docid=93%
146}}.

\bibitem[{\citenamefont{Klimenko and Mitselmakher}(2004)}]{waveburst}
\bibinfo{author}{\bibfnamefont{S.}~\bibnamefont{Klimenko}} \bibnamefont{and}
  \bibinfo{author}{\bibfnamefont{G.}~\bibnamefont{Mitselmakher}},
  \bibinfo{journal}{Classical Quantum Gravity} \textbf{\bibinfo{volume}{21}},
  \bibinfo{pages}{S1819} (\bibinfo{year}{2004}).

\bibitem[{\citenamefont{Sylvestre}(2002)}]{sylvestre}
\bibinfo{author}{\bibfnamefont{J.}~\bibnamefont{Sylvestre}},
  \bibinfo{journal}{Phys. Rev. D} \textbf{\bibinfo{volume}{66}},
  \bibinfo{pages}{102004} (\bibinfo{year}{2002}).

\bibitem[{\citenamefont{Prestegard et~al.}(2012)\citenamefont{Prestegard,
  Thrane et~al.}}]{stamp_glitch}
\bibinfo{author}{\bibfnamefont{T.}~\bibnamefont{Prestegard}},
  \bibinfo{author}{\bibfnamefont{E.}~\bibnamefont{Thrane}},
  \bibnamefont{et~al.}, \bibinfo{journal}{Classical Quantum Gravity}
  \textbf{\bibinfo{volume}{28}}, \bibinfo{pages}{095018}
  (\bibinfo{year}{2012}).

\bibitem[{\citenamefont{Abbott et~al.}(2009)}]{ligo_sgr}
\bibinfo{author}{\bibfnamefont{B.}~\bibnamefont{Abbott}} \bibnamefont{et~al.}
  (\bibinfo{collaboration}{LIGO Scientific Collaboration}),
  \bibinfo{journal}{Astrophys. J. Lett.} \textbf{\bibinfo{volume}{701}},
  \bibinfo{pages}{68} (\bibinfo{year}{2009}).

\bibitem[{\citenamefont{Farin}(1996)}]{bezier}
\bibinfo{author}{\bibfnamefont{G.}~\bibnamefont{Farin}},
  \emph{\bibinfo{title}{Curves and Surfaces for CAGD, Fourth Edition: A
  Practical Guide}} (\bibinfo{publisher}{Academic Press},
  \bibinfo{year}{1996}).

\bibitem[{\citenamefont{{Harry, G. M. for the LIGO Scientific
  Collaboration}}(2010)}]{aligo}
\bibinfo{author}{\bibnamefont{{Harry, G. M. for the LIGO Scientific
  Collaboration}}}, \bibinfo{journal}{Classical Quantum Gravity}
  \textbf{\bibinfo{volume}{27}}, \bibinfo{pages}{084006}
  (\bibinfo{year}{2010}).

\bibitem[{\citenamefont{Santamar\'ia and Ott}(2011)}]{lucia}
\bibinfo{author}{\bibfnamefont{L.}~\bibnamefont{Santamar\'ia}}
  \bibnamefont{and} \bibinfo{author}{\bibfnamefont{C.~D.} \bibnamefont{Ott}},
  \bibinfo{journal}{LIGO DCC} p. \bibinfo{pages}{T1100093}
  (\bibinfo{year}{2011}),
  \bibinfo{note}{\url{https://dcc.ligo.org/LIGO-T1100093-v2/public}}.

\bibitem[{\citenamefont{Abbott et~al.}(2010)}]{grb:s5vsr1}
\bibinfo{author}{\bibfnamefont{B.~P.} \bibnamefont{Abbott}}
  \bibnamefont{et~al.}, \bibinfo{journal}{Astrophys. J.}
  \textbf{\bibinfo{volume}{715}}, \bibinfo{pages}{1438} (\bibinfo{year}{2010}).

\bibitem[{\citenamefont{{Acernese, F. for the Virgo
  Collaboration}}(2006)}]{virgo}
\bibinfo{author}{\bibnamefont{{Acernese, F. for the Virgo Collaboration}}},
  \bibinfo{journal}{Classical Quantum Gravity} \textbf{\bibinfo{volume}{23}},
  \bibinfo{pages}{S63} (\bibinfo{year}{2006}).

\bibitem[{\citenamefont{Blackburn et~al.}(2008)}]{lsc_glitch}
\bibinfo{author}{\bibfnamefont{L.}~\bibnamefont{Blackburn}}
  \bibnamefont{et~al.}, \bibinfo{journal}{Classical Quantum Gravity}
  \textbf{\bibinfo{volume}{25}}, \bibinfo{pages}{184004}
  (\bibinfo{year}{2008}).

\bibitem[{\citenamefont{Cornish and Romano}(2013)}]{cornish_romano}
\bibinfo{author}{\bibfnamefont{N.~J.} \bibnamefont{Cornish}} \bibnamefont{and}
  \bibinfo{author}{\bibfnamefont{J.~D.} \bibnamefont{Romano}},
  \bibinfo{journal}{Phys. Rev. D} \textbf{\bibinfo{volume}{87}},
  \bibinfo{pages}{122003} (\bibinfo{year}{2013}).

\bibitem[{\citenamefont{Anderson et~al.}(2001)\citenamefont{Anderson, Brady,
  Creighton, and Flanagan}}]{box}
\bibinfo{author}{\bibfnamefont{W.~G.} \bibnamefont{Anderson}},
  \bibinfo{author}{\bibfnamefont{P.~R.} \bibnamefont{Brady}},
  \bibinfo{author}{\bibfnamefont{J.~D.~E.} \bibnamefont{Creighton}},
  \bibnamefont{and} \bibinfo{author}{\bibfnamefont{{\'E}.~{\'E}.}
  \bibnamefont{Flanagan}}, \bibinfo{journal}{Phys. Rev. D}
  \textbf{\bibinfo{volume}{63}}, \bibinfo{pages}{042003}
  (\bibinfo{year}{2001}).

\bibitem[{\citenamefont{Murphy et~al.}(2013)\citenamefont{Murphy, Tse, Raffai
  et~al.}}]{murphy_sgr}
\bibinfo{author}{\bibfnamefont{D.}~\bibnamefont{Murphy}},
  \bibinfo{author}{\bibfnamefont{M.}~\bibnamefont{Tse}},
  \bibinfo{author}{\bibfnamefont{P.}~\bibnamefont{Raffai}},
  \bibnamefont{et~al.}, \bibinfo{journal}{Phys. Rev. D}
  \textbf{\bibinfo{volume}{87}}, \bibinfo{pages}{103008}
  (\bibinfo{year}{2013}).

\end{thebibliography}


\end{document}